# Solar Term Anomaly in China Stock Market: Evidence from Shanghai Index

Zhou Tianbao[a,c], Li Xinghao[b], Zhao Junguang[a,*]

[a] *College of Science, Beijing Forestry University, 100083, Beijing China*
[b] *School of Information Science & Technology, Beijing Forestry University, 100083, Beijing China*
[c] *School of Economics and Management, Beihang University, 100191, Beijing China*

**Authors' e-mail:**
1. Mr. Zhou Tianbao[a,c]: michaelzhou@bjfu.edu.cn, michaelzhou@buaa,edu.cn
2. Mr. Li Xinghao[b]: lixinghao@bjfu.edu.cn
3. Dr. Zhao Junguang[a]*: zhaojg@bjfu.edu.cn

**Abstract**: This paper investigates the solar term effect (anomaly) in China stock market as a supplementary to the existing literature of calender effect. Based on a regression framework, this paper verifies the existence of solar term effect in Shanghai Index in multiple dimensions: inter-solar-term analysis, full sample analysis at mean level and risk level as well as the turn of solar term effect. Several solar terms have been found to cause significant positive and negative value to the return such as solar term 1,3 and 4. and bring high volatility such as solar term 8, 11 and 14. The result is reliable and robust under the Extreme Bound Analysis and various assumptions of error's distribution in IGARCH model. These findings give readers a new perspective to view calender effect under the influence of traditional Chinese culture that solar terms affect the market through affecting investors' mood, expectation, enthusiasm, etc. which is a good evidence to the "Culture bonus hypothesis" proposed by Chen and Chien (2011) and the possible influence by the Chinese culture in other Asian markets (Yuan and Gupta, 2014).

# 1. Introduction

There has been continuous voice in the efficiency and regulation of stock market since Fama (1970) proposes the Efficient Market Hypothesis (or EMH). Fama (1970) suggests that all information is fully and rapidly reflected in the stock prices and investors are not able to apply any strategy (e.g. technical and fundamental analysis) to win excess return based on the historical data of stocks. In other words, stocks are unpredictable with little regulation. However, since 1980s, behavioral financial economist and scholars has found many anomalies in stock markets that challenge the EMH, among which turns out to be the famous one: Calender anomaly (or Calender effect). Calendar effect, first discovered in the study of stock returns, refers to a date-related anomaly in which returns are significantly higher or lower than the average at a particular time and investors can benefit through their anticipation on a certain day based on the historical regulation. Prevailing calender effects include weekend effect, month effect, holiday effect, turn of the month effect, etc. Thaler (1987) and Schwert (2003) give readers summaries of abundant introductions of anomalies in the early time. It can be seen that calender effect is based on various divisions of time and date specifications (e.g. time spot, certain period of time, seasons, etc.) and that concept is independent from the traditional financial and economic factors which gives us another perspective to seek the regulation of stock market.

In recent years, calender anomalies in China stock market has drawn much attention and especially, a number of studies discover calender anomaly based on Chinese lunar calender (or call Chinese farmer calender) suck as lunar new year effect (Teng and Yang, 2018). These findings are also investigated in many other Asian stock markets (McGuinness and Harris, 2011; Yang, 2016). Given that the traditional Chinese twenty-four solar terms is a refined division of annual season and climate, solar term effect (anomaly) is definitely a notable factor in China stock market. In fact, the chronology of Chinese twenty-four solar terms is also based on the lunar calender and it is more suitable for studying date and time with traditional Chinese characteristics. The date of twenty-four solar terms is consist with some of what Gann (2014) mention in his book[1]. Therefore, solar terms are not only discovered in China market but more importantly, early western scholars may also notice that point on their own. In brief, there are 24 refined seasons within a year under the Chinese twenty-four solar terms instead of what we are using internationally today (i.e. only spring, summer, fall and winter). For the complete introduction of the Chinese twenty-four solar terms, please see in Appendix A. The date (chronology) of Chinese twenty-four solar terms is presented in Table 0.

**Table 0.** Date of Chinese twenty-four solar terms

| Solar term order | Solar term name (in pinyin) | Solar term date | Lunar season | Solar term order | Solar term name (in pinyin) | Solar term date | Lunar season |
|---|---|---|---|---|---|---|---|
| 1 | Xiaohan | Jan 5-Jan 7 | Winter | 13 | Xiaoshu | Jul 6-Jun 8 | Summer |
| 2 | Dahan | Jan 20-Jan 21 | | 14 | Dashu | Jul 22- Jul24 | |

---

[1] Besides Gann's Wheel, Gann's line, etc. Gann also points out a series of date on which stock is mutable. Those date mostly comes from dividing a year into 4, 6, 8, 12, etc. equal parts which is exactly the same as how Chinese twenty-four solar terms originate. In his book, Gann believe time and date could be a independent factor affecting stocks but he just does not associate it with the division of Chinese twenty-four solar terms.

| | | | | | | | |
|---|---|---|---|---|---|---|---|
| 3 | Lichun | Feb 3-Feb 5 | | 15 | Liqiu | Aug7-Aug 9 | |
| 4 | Yushui | Feb 18-Feb 20 | | 16 | Chushu | Aug 22- Aug 24 | |
| 5 | Jingzhe | Mar 5-Mar 7 | Spring | 17 | Bailu | Sept 7-Sept 9 | Fall |
| 6 | Chunfen | Mar 20-Mar 22 | | 18 | Qiufen | Sept 22- Sept 24 | |
| 7 | Qingming | Apr 4- Apr 6 | | 19 | Hanlu | Oct 8-Oct 9 | |
| 8 | Gu'yu | Apr 19-Apr 21 | | 20 | Shuangjiang | Oct 23-Oct 24 | |
| 9 | Lixia | May 5-May 7 | | 21 | Lidong | Nov 7-Nov 8 | |
| 10 | Xiaoman | Mar 20-May 22 | Summer | 22 | Xiaoxue | Nov 22-Nov 23 | Winter |
| 11 | Mangzhong | Jun 5- Jun 7 | | 23 | Daxue | Dec 6-Dec 8 | |
| 12 | Xiazhi | Jun 21-Jun 22 | | 24 | Dongzhi | Dec 20-Dec 21 | |

Note we present the twenty-four solar terms in an order of international solar calender. In fact solar term *3* (Lichun) is the first one according to Chinese tradition as spring is regarded as the beginning of the year. Since the solar terms are developed by lunar calender, the date is not fixed in solar calender. Therefore, the 3$^{rd}$ and 7$^{th}$ column in the table provide a range of when the solar term will fall. Each solar term has its unique meaning according to the Chinese name but for the convenience, we only say the order in the following sections. Especially, Chinese lunar new year (i.e. Spring Festival) happens no earlier than solar term *2* (Dahan) and no later than solar term *4* (Yushui).

As a matter of fact, studies of solar term effect in China stock market are rare since they are mostly published only in some Chinese platforms and the gap of this research still remains large. Ni (2013) explores solar term effect in Shanghai Index and Shenzhen Index in early days finding that there is a positive effect on return on Dahan in Shanghai Index and Shenzhen Index and Dongzhi is also positive if loosing the significance level to 10% confirming the existence of solar term effect in China stock market. He also concludes that not all sector indexes show obvious solar term effect, but for those sectors occupying a big proportion in China stock market all shows positive effect on Dahan. The drawback of his study as he also says, is the limit of data which is only from 2000 to 2012. Wang (2017) takes a similar study with a larger amount of data (1997-2016) of Shanghai-Shenzhen 300 Index. She finds a positive return on Chunfen and Lichun which is much more consistent with what we find in this article. Whether solar terms effect can be explained by Holiday Effect and Festival Effect was also tested in several other markets of China and countries abroad. On the other hand, Zhang, Ou and Xu (2018) who study the relationship between solar term effect and holiday effect finding out that there is a positive return during the Spring Festival. Qingming and Dongzhi. Worth mentioning is that our team (Zhou, Li and Wang, 2021) conducts a brand-new study on the relationship between stock index turning points (reversals) and Chinese twenty-four solar terms finding that overall the trend of stock indices is more likely to reverse around solar terms.

Generally speaking, current studies of solar terms are limited and lack of systematicness.to sufficient solar term effect verification . This paper studies solar term effect in Shanghai Index based on a regression framework using a large span of sample from 1995 to 2022. The main contribution of this paper to the literature is that we verify solar term effect from multiple dimensions: inter-solar-term, full sample mean level and full sample risk level and by various robust tests. We systematically analyze abundant categories of possible solar term effect in a large span of sample and we further explain the relationship between existing calender effects, traditional Chinese culture and solar term effect in an effort to figure out the potential factors that shape the solar term effect in China stock market.

In inter-solar-term analysis, apart from using purely dummy regression model, we also introduce the Extreme Bound Analysis (EBA) for more reliable robust test since that method is more stringent. We find that several solar terms at the beginning of a year have significant positive and negative return such as solar term *1*, *3* and *4*. Some of them passed the EBA robust test. In full sample analysis, we use AR (1)-IGARCH (1,1) model to capture return feature at mean level and risk level correspondingly. Involving with non-solar-term days (henceforth, normal days) and solar-term days, we discover remarkable features of a couple of solar terms having significant value and contribution based on multiple error assumptions in an effort to verify the strongness (or called efficiency) of each solar term. The result of turn of solar term effect is also notable. Those significant solar terms in inter-solar-term analysis and full sample analysis are highly consistent (mostly are early in the year such as solar term 1, 2, 3 and 4 ). Therefore, the existing solar term effect is confirmed to be robust and stable in all cases under multi-dimensional analysis. The result reveals the potential impact of Chinese culture and investors mood as a sort of "Early year effect" because the Spring Festival and Spring solar terms happen at that period. This interesting result also support the previous studies of Month effect and January effect as the first two solar terms are in January. It also verify the impact of existing Chinese lunar new year (Spring Festival) effect as solar term *3* (Lichun) is always closely before and after Spring Festival. This can be explained by the "Culture bonus hypothesis" proposed by Chen and Chien (2011). The solar term effect is a new perspective to understand the influence that Chinese investors get more hopes and expectation since a fresh new year as well as the active trading and impetuousness by the rising temperature, revival of lives in warm seasons which is a inrooted influence of Chinese people.

To sum up, this paper answers 1): which solar terms are significant in inter-solar-term comparison. 2): which solar terms are significant out of full sample both at mean level and risk (volatility) level. 3) the turn of solar term effect at risk level. 4) the logic of how solar term effect may shape in China stock market. The remainder of this paper is as follows: Section 2 introduces our data and methodology; Section 3 presents the empirical result; Section 4 is the conclusions and Section 5 is the discussions.

## 2. Data and methodology

The data we used in this study is the daily return of Shanghai Index[2] in China (code: 000001) from the first trading day in1995 (Jan 1, 1995) to the last trading day in 2022 (Dec 30, 2022). That's 28 entire years and there are totally 24×28=672 solar terms in our dataset (some of the solar terms are not on trading days, so we exclude them). Given that the daily price series of Shanghai Index is not stationary, we simply study on its daily return series through out this paper. The data is easy and free to obtain in most securities website[3].

---

[2] Shanghai Index (Shanghai securities composite index) is the earliest and the most influential stock index in China. It is a weighted composite index covering a major part of blue-chip stocks and dominant sectors in China market such as banks, securities and insurance, compunctions, state-owned enterprise, energy, etc. As it contains thousands of stocks, it is relatively smooth and steady in trend comparing with other technology index. Usually, the whole market is going up when Shanghai Index has a significant rise, Shanghai Index is used as a wind indicator to the overall heat and sentiment of the market . Thus, Shanghai Index is suitable to be a proxy to China stock market.

A predominant and reasonable way to examine calender effect is to apply the regression model with several dummy variables corresponding to the days we refer. Note that the study on solar terms effect (anomaly) is slightly differently to those on weekend effect or month effect. Monday to Friday occupy all trading days in a week and January to December cover all the trading months within a year. However, a trading day is not necessarily to be a solar term day or another. Solar terms follow Chinese lunar calender so we need to distinguish solar-term days and non-solar-term days before making up the model.

*2.1. Inter-solar-term analysis*

In this section, we solely select solar-term days from our sample and study purely on the feature and comparison among twenty-four solar terms. In such background, daily returns have the following dummy regression model:

$$R_t = \mu + \sum_{i=1}^{24} \alpha_i ST_{it} + \varepsilon_t \qquad (1)$$

where $R_t$ denotes the daily return, $ST_{it}$ denotes $i^{th}$ solar term dummy variable such that $ST_{it} = 1$ if solar term No.*i* falls on that day, otherwise $ST_{it} = 0$.

It is common sense that equation (1) cannot be estimated due to perfect collinearity of the full use of dummy variables which is called dummy variable trap (i.e. $\sum_{i=1}^{24} ST_{it} = 1$). This means that in a regression model with intercept, the intercept and 24 terms are linearly related. In other words, after setting 24 dummy variable for 24 solar terms correspondingly and when all dummy variable equal to zero, there is no a $25^{th}$ solar term for the intercept to be explained. In order to solve the problem, some restrictions are needed to be claimed on how to specify the dummy variables. A widely-used approach is the dummy code by deleting one dummy variable from all. In other words, we need to specify *n-1* dummy variables if there are *n* cases in total. In this study, we finally specify 23 dummy variables as there are 24 solar terms in total. The unselected (deleted) one is called reference variable (reference solar term) which servers as the constant terms in the regression. Therefore, equation (1) is modified to:

$$R_t = \mu + \sum_{i \neq st*} \alpha_i ST_{it} + \varepsilon_t \qquad (2)$$

where *st\** is the reference solar term. The feature of *st\** is reflected on $\mu$, $\alpha_i$ is the coefficient estimation of solar term *i*. We can conclude from equation (2) that:

$$E(R \mid ST = st^*) = \mu \qquad (3)$$
$$E(R \mid ST = i) = \mu + \alpha_i \qquad (4)$$

$\alpha_i$ is the contribution on return compared with the reference solar term. Thus, $\alpha_i$ is on a relative concept whereas $\mu$ is the absolute contribution to return. Now, another issue is raised, how do we choose the reference term? Theoretically, the choice of picking the reference term is arbitrary, but we simplify it and make the choice more efficient according to the statistics feature of solar terms which we analyze in details in the following charter.

Worth mentioning is that such dummy variable will not exist in a regression model without intercept as equation (1'):

---
[3] The data is open and free to download from securities websites like https://www.eastmoney.com and http://data.10jqka.com.cn/
Or from leading dataset website: https://www.gtarsc.com/, and WIND database in China.

$$R_t = \sum_{i=1}^{24} \alpha_i ST_{it} + \varepsilon_t \qquad (1')$$

where none of the dummy variables is required to be excluded from the model and picking a reference term is also no longer needed. However, predominant regression models always contain a intercept as the estimation for R-square and interpretation for coefficients may not be accessible in a model without intercept, causing extra problems, see Gerald (1977), Jobson (1982) and Hawkins (1980) for further studies. Many statistics software always introduce a intercept by default.

In addition, another way to examine the robustness of estimated coefficient in a regression is the Extreme Bound Analysis or EBA (Leamer, 1983, 1985). EBA suggests a new way to assess the error of coefficient in a panel data model since heteroskedasticity leads to consistent but inefficient least squares estimates in a linear regression model. EBA may result in high error for coefficients and the coefficient is regarded as robust under EBA when it falls within the upper and lower bound of EBA without changing its sign, otherwise, it is fragile. EBA is regarded as a very stringent criterion for robustness (Salai-Martin, 1997). As a result, in applications it is generally the case that most (if not all) of the candidate regression variables are declared fragile. Hence, if a variable is regarded as robust under EBA, it is quite likely to be cataloged as robust with other methods dealing with model uncertainty.

In a traditional regression model $y = X\beta + \varepsilon$, EBA focuses on estimating the covariance matrix for the coefficient in $\beta$ or to say more exactly, the error of each estimated parameter. Finally, we have the lower and upper bound of each coefficient $\acute{\beta}_i$ which are $\acute{\beta}_i - u\sigma_i$ and $\acute{\beta}_i + u\sigma_i$.

Each $\sigma_i$ is calculated from the EBA covariance matrix. There are a series of estimators for it but the most recommended is the *HC3* estimator (Winkelried and Iberico, 2018). For the detail of the rest estimators (e.g. *HC1* and *HC2*), see Francisco et al. (2006) and MacKinnon (1985). Winkelried and Iberico (2018) apply EBA method on weekend effect by simplifying *HC3* estimator into a more clear and concise form which has identical results as the original one in our study. The *HC3* covariance matrix Winkelried and Iberico (2018) adapt is constructed as follow:

$$HC3 = (X'X)^{-1}X'AX(X'X)^{-1} \qquad (5)$$

where $A = diag(a_i)$ and

$$a_i = \frac{e_t^2}{(1-h_t)^2} \qquad (6)$$

$h_1, h_2, \ldots, h_n$ is the diagonal elements of $X'(X'X)^{-1}X$. By squaring root of the diagonal elements of *HC3*, we can finally obtain the standard error of each estimated coefficient under EBA. Therefore, EBA plays as a supplementary to test the robustness of dummy regression.

*2.2. Full sample analysis*

The analysis in Section 2.1 only focuses among 24 solar terms but not considering other non-solar term days (henceforth "normal day"). We are going to extend it to full sample range in this section. Based on previous works, we know that financial series always accompany with auto-correlation and heteroscedasticity. To address the problem above, many papers combine time series framework with dummy regression (e.g. Halil and Berument, 2003; Sabri et al., 2017). We thus include a AR(1) process to examine the solar term anomaly in the return (mean level). The

process is as follow:

$$R_t = \mu + \sum_{i=1}^{24} \alpha_i ST_{it} + rR_{t-1} + \varepsilon_t \tag{7}$$

where the symbols remain the same as we mention in section 2.1, $R_{t-1}$ denotes the return of last trading day, $r$ is its coefficient. In equation (7), there are 25 types of day: 24 solar term days plus normal day. Therefore, it will not cause dummy variable trap by specify all 24 solar terms. When all dummy variables equal to zero, it refers to normal day. The advantage is that it make it possible for all solar terms to be interpreted equally and the constant $\mu$ contains the information of normal days. The lagged term of return also contributes to the accuracy of the model.

Besides, we also want to examine the solar term anomaly at the risk level (i.e. variance or volatility). Therefore, we allow variances of errors to be time varying in equation (7) through including a conditional heteroscedasticity equation that captures the risk. Well, the error terms now have a mean of zero and a time varying variance of $h_t^2$. Engle (1982) proposes the ARCH ($q$) model firstly that allows the squared error to be affected by its lagged terms. Many other derivations of ARCH model are further developed by scholars and researchers (Baillie and DeGennaro, 1990). It can be easily verified that the sample of Shanghai Index is more likely to meet the condition of IGARCH (1,1) model[4] instead of the traditional GARCH (1,1) model. We make the variance model as follow:

$$h_t^2 = \sum_{i=1}^{24} \alpha_i ST_{it} + \gamma \varepsilon_{t-1}^2 + \beta h_{t-1}^2 \tag{8}$$

where IGARCH (1,1) requests that $\gamma + \beta = 1$. Note that we do not know the actual distribution of the error terms so we estimate the IGARCH equation under multiple distribution assumptions in order to test the robustness of solar terms as the ordinary Normal Distribution assumption (i.e. $\varepsilon_t \sim N(0, h_t^2)$) for error terms in financial series may be too loose. Therefore, two more distribution assumptions are considered in the estimation: Student-$t$ distribution and Generalized error distribution (GED)[5]. Nelson (1991) proposes to take GED to capture the fat tails feature observed in the distribution of financial time series. Brooks and Persand (2001) and Doyle and Chen (2009) analyze the sensitivity to the choice of error distribution assumptions.

## 3. Empirical results

### 3.1. Data statistics and inter-solar-term result

Table 1 shows the statistics of all solar term days' return:

**Table 1 Statistics of solar term days return**

| Order | Mean value | Standard deviation | Skewness (normal is 0) | Kurtosis (normal is 3) | Shapiro Normality |
|---|---|---|---|---|---|

---

[4] The parameters γ plus β is so close to unity. It is reasonable to say that there is a unit root in that model, thus we use IGARCH (1,1) instead of traditional GARCH(1,1). Relative works see Thomas and Stărică, 2004; Ling, 2007.

[5] $\Gamma( )$ is the gamma function, $v > 0$, $\lambda$ is the constant. $\lambda = \sqrt{\frac{2^{-2/v} \Gamma(1/v)}{\Gamma(3/v)}}$. $v$ equal 2 and $\lambda$ equal 1, it turns out to be standard Normal Distribution. See Hamilton (1994) for details.

|    | Mean value | *t*-test<br>(*p*-value) |        |         |        | test<br>(*p*-value) |
|----|------------|-------------------------|--------|---------|--------|---------------------|
| 1  | 0.0086***  | 3.0405<br>(0.0067)      | 0.0127 | 0.7613  | 3.1868 | 0.9376<br>(0.2167)  |
| 2  | -0.0049    | -1.3098<br>(0.2067)     | 0.0165 | -1.2881 | 4.5201 | 0.8795**<br>(0.0210) |
| 3  | 0.0124*    | 1.8521<br>(0.0887)      | 0.0242 | 1.7493  | 6.0738 | 0.7901***<br>(0.0050) |
| 4  | -0.0083    | -1.0349<br>(0.3211)     | 0.0291 | -1.9032 | 5.7661 | 0.7332***<br>(0.0012) |
| 5  | 0.0043     | 1.0003<br>(0.3297)      | 0.0193 | 1.7045  | 6.9008 | 0.8475***<br>(0.0048) |
| 6  | 0.0057*    | 1.7614<br>(0.095)       | 0.0141 | -0.9759 | 5.0036 | 0.8895**<br>(0.0315) |
| 7  | -0.0014    | -0.3504<br>(0.7351)     | 0.0120 | -0.6166 | 1.8963 | 0.8718<br>(0.1287)  |
| 8  | 0.0020     | 0.6156<br>(0.5458)      | 0.0144 | 0.8935  | 3.4239 | 0.9271<br>(0.1535） |
| 9  | 0.0001     | 0.0308<br>（0.9759）    | 0.0215 | -0.9663 | 5.1959 | 0.8771**<br>（0.0652） |
| 10 | 0.0032     | 0.7681<br>（0.4519）    | 0.0191 | -1.6153 | 6.6511 | 0.8505***<br>（0.0054） |
| 11 | 0.0039     | 0.8759<br>（0.3926）    | 0.0195 | 2.4536  | 8.3173 | 0.6461***<br>（0.0000） |
| 12 | 0.0010     | 0.3147<br>( 0.7564)     | 0.0145 | 0.0174  | 3.5854 | 0.9528<br>(0.4129)  |
| 13 | 0.0065     | 1.1873<br>(0.2497)      | 0.0245 | 0.1039  | 3.8640 | 0.8955**<br>(0.0340 |
| 14 | 0.0058*    | 1.7290*<br>(0.1000)     | 0.0152 | 0.2389  | 3.3244 | 0.9529<br>(0.4145)  |
| 15 | -0.0020    | -0.5491<br>(0.5897)     | 0.0158 | -0.3204 | 3.1660 | 0.9699<br>(0.7760)  |
| 16 | -0.0002    | -0.1465<br>(0.8850)     | 0.0088 | -0.0565 | 2.5356 | 0.9700<br>(0.7564)  |
| 17 | 0.0028     | 0.8128<br>(0.4269)      | 0.0153 | 0.3173  | 3.3188 | 0.9693<br>(0.7635)  |
| 18 | 0.0033     | 0.5064<br>(0.6187)      | 0.0288 | 0.8272  | 4.5370 | 0.8912**<br>( 0.0338) |
| 19 | 0.0023     | 0.4836<br>(0.6348)      | 0.0201 | -0.3708 | 2.2837 | 0.9585<br>(0.5733)  |
| 20 | 0.0048     | 0.8268<br>(0.4186)      | 0.0260 | 2.4612  | 9.4867 | 0.7246***<br>(0.0000) |
| 21 | -0.0037    | -0.9698<br>(0.3443)     | 0.0172 | -1.6006 | 4.8712 | 0.7819***<br>(0.0004) |
|    | Mean value | *t*-test<br>(*p*-value) |        |         |        | test<br>(*p*-value) |
| 22 | 0.0005 | 0.2834 | 0.0077 | -0.6819 | 3.1168 | 0.9444 |
| | | (0.7801) | | | | (0.3162) |
| 23 | 0.0017 | 1.0396 | 0.0072 | -0.0264 | 2.5273 | 0.9814 |
| | | (0.3123) | | | | (0.9576) |
| 24 | 0.0029 | 1.0525 | 0.0123 | 0.5661 | 4.0925 | 0.9267 |
| | | (0.3065) | | | | (0.1510) |

Note: Test for mean :alternative hypothesis is that true mean is not equal to 0; Normality test: alternative hypothesis is that the distribution is not normal.

* stands for 10% level significance, ** stands for 5% level significance, *** stands for 1% level significance.

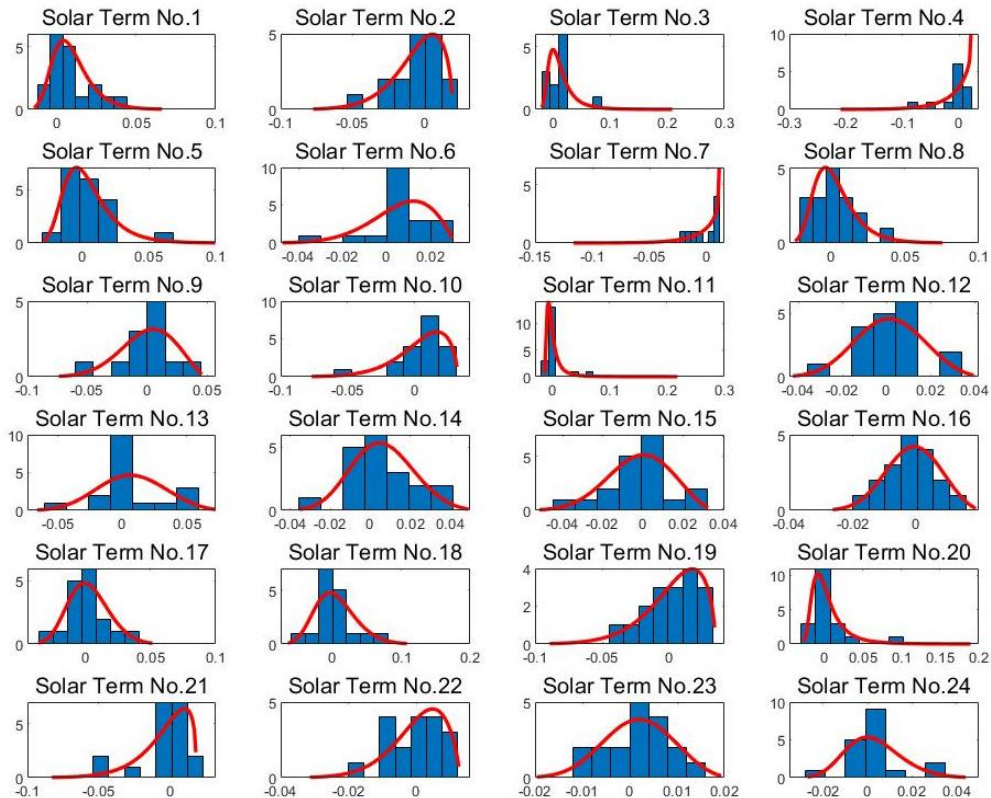

**Fig. 1.** Broad view of sample's distribution (with Kernel density) of twenty-four solar terms' return

According to Table 1 and Figure 1, most solar terms have a mean value very close to zero with a high kurtosis over 3 (3 is the kurtosis of Normal distribution). the distribution of returns in our sample show tick tail and high peak. Also, a part of them show a high skewness which is good evidence for the asymmetry of solar term analysis. Solar term 1, 3, 6, and 14 are significant in mean value and solar term 2 and 4 are negative and relatively high. Solar term 2 to 6, 8, 10, 11, 13, 18 ,20 and 21 do not follow Normal distribution significantly. These solar terms have special features out of all 24 terms being the potential candidate for further attention. However, the drawback is that these preliminary tests above are based on Normal distribution assumption, thus we need further examine carefully. Given that, it is important to realize that not all solar term have special and significant effect on returns and there may be only a part of them useful in the end which we call efficient solar terms. The rest solar terms are mainly accompanied by randomness and less featured which we call inefficient solar terms.

We then implement the dummy regression equation (2) in Section 2.1 and finally get four groups of result having significant reference terms, they are solar term 1, 3, 4 and 13. Table 2 presents the overall estimation of coefficient in each group. In Panel A and Panel B, we find that solar term 1 and 3 are significant in contributing high positive values to daily return (the dependent variable, at 5% level) while other solar terms in such panels do not have significant positive return on their own but showing relatively negative effects to solar term 1 and 3 (solar term 2, 4, 15 etc.). We can regard them as the opposite solar terms correspondingly. In Panel C and Panel D, solar term 4 shows a significant negative effect to the overall return whereas solar term 2 does not pass the significance test but also having a significance level closely. The rest of the solar terms on the other hand display a relatively positive effect to solar term 2 and 4. In Panel E, solar term 13 is the reference term having a significance at 10%. It is obvious that no matter talking about reference terms or others, those solar terms showing up in Table 2 are what always show up. In other words, efficient solar terms are almost fixed in whatever panels. Besides, There is no linear explanation between dummy variables as Tolerance and VIF all confirm that there's no multi-collinearity in each panel.[6]

Table 2   Inter-solar-term regression result

| Solar Term | Coefficient (Std error) | Lower interval (95%) | Upper interval (95%) | Tolerance | VIF |
|---|---|---|---|---|---|
| *Panel A* | | | | | |
| No.1 (Reference term) | 0.0086** (0.0041) | 0.0006 | 0.0166 | - | - |
| No.2 | -0.0136** (0.0058) | -0.0250 | -0.0021 | 0.5361 | 1.8650 |
| No.4 | -0.0169*** (0.0065) | -0.0297 | -0.0042 | 0.6246 | 1.6008 |
| No.15 | -0.0106* (0.0058) | -0.0220 | 0.0007 | 0.5361 | 1.8650 |
| No.21 | -0.0123** (0.0057) | -0.0236 | -0.0011 | 0.5240 | 1.9082 |
| *Panel B* | | | | | |
| No.3 (Reference term) | 0.0124** (0.0050) | 0.0025 | 0.0223 | - | - |
| No.2 | -0.0174*** (-0.0065) | -0.0302 | -0.0045 | 0.4247 | 2.3542 |
| No.4 | -0.0208*** (0.0071) | -0.0347 | -0.0068 | 0.5153 | 1.9403 |
| No.7 | -0.0138* (0.0079) | -0.0293 | 0.0016 | 0.6033 | 1.6573 |
| No.9 | -0.0122* (0.0065) | -0.0262 | 0.0017 | 0.5153 | 1.9403 |
| No.12 | -0.0114* (0.0065) | -0.0241 | 0.0012 | 0.4128 | 2.4220 |

---

[6] Tolerance and VIF are indices showing the degree of multi-collinearity. VIF is the reciprocal of Tolerance. When VIF is close to 1, it means there's no multi-collinearity. VIF=10 is often used as a threshold to judge whether there is multi-collinearity in the model.

| | | | | | |
|---|---|---|---|---|---|
| No.15 | -0.0144** | -0.0272 | -0.0016 | 0.4247 | 2.3542 |
| | (0.0065) | | | | |
| No.16 | -0.0127** | -0.0254 | -0.0000 | 0.4128 | 2.4220 |
| | (0.0065) | | | | |
| No.21 | -0.0162** | -0.0289 | -0.0034 | 0.4128 | 2.4220 |
| | (0.0065) | | | | |
| *Panel C* | | | | | |
| No.2 | -0.0050 | -0.0131 | 0.0032 | - | - |
| (Reference term) | (0.0042) | | | | |
| No.1 | 0.0136** | 0.0021 | 0.0250 | 0.5106 | 1.9584 |
| | (0.0085) | | | | |
| No.3 | 0.0174*** | 0.0045 | 0.0302 | 0.6119 | 1.6339 |
| | (0.0065) | | | | |
| No.6 | 0.0106* | -0.0008 | 0.0222 | 0.5227 | 1.9128 |
| | (0.0065) | | | | |
| No.13 | 0.0115*** | 0.0000 | 0.0229 | 0.5106 | 1.9584 |
| | (0.0058) | | | | |
| No.14 | 0.0108* | -0.0005 | 0.0222 | 0.5106 | 1.9584 |
| | (0.0058) | | | | |
| No.20 | 0.0097* | -0.0016 | 0.0212 | 0.5106 | 1.9584 |
| | (0.0058) | | | | |
| *Panel D* | | | | | |
| No.4 | -0.0083* | -0.0182 | 0.0015 | - | - |
| (Reference term) | (0.0050) | | | | |
| No.1 | 0.0169*** | 0.0042 | 0.0297 | 0.4128 | 2.4220 |
| | (0.0065) | | | | |
| No.3 | 0.0208*** | 0.0068 | 0.0347 | 0.5153 | 1.9403 |
| | (0.0071) | | | | |
| No.5 | 0.0126* | -0.0000 | 0.0253 | 0.4128 | 2.4220 |
| | (0.0065) | | | | |
| No.6 | 0.0140** | 0.0012 | 0.0269 | 0.4247 | 2.3542 |
| | (0.0065) | | | | |
| No.10 | 0.0116* | -0.0010 | 0.0243 | 0.4128 | 2.4220 |
| | (0.0065) | | | | |
| No.11 | 0.0122* | -0.0005 | 0.0251 | 0.4247 | 2.3542 |
| | (0.0065) | | | | |
| No.13 | 0.0148** | 0.0021 | 0.0275 | 0.4128 | 2.4220 |
| | (0.0065) | | | | |
| No.14 | 0.0142** | 0.0015 | 0.0269 | 0.4128 | 2.4220 |
| | (0.0065) | | | | |
| No.17 | 0.0112* | -0.0016 | 0.0240 | 0.4247 | 2.3542 |
| | (0.0065) | | | | |
| No.18 | 0.0117* | -0.0011 | 0.0245 | 0.4247 | 2.3542 |
| | (0.0065) | | | | |
| No.20 | 0.0131** | 0.0004 | 0.0258 | 0.4128 | 2.4220 |
| | (0.0065) | | | | |
| No.21 | 0.0113* | -0.0015 | 0.0241 | 0.4247 | 2.3542 |
| | (0.0065) | | | | |
| *Panel E* | | | | | |
| No.13 (Reference term) | 0.0065* (0.0041) | -0.0014 | 0.0145 | - | - |
| No.2 | -0.0115** (0.0058) | -0.0229 | -0.0001 | 0.5362 | 1.865 |
| No.4 | -0.0149** (0.0065) | -0.0276 | -0.0022 | 0.6247 | 1.6008 |
| No.21 | -0.0103* (0.0057) | -0.0216 | 0.001 | 0.524 | 1.9083 |
| obs | | | 436 | | |

\* stands for 10% level significance, \*\* stands for 5% level significance, \*\*\* stands for 1% level significance.

Table 3 shows the interval of each reference term under EBA as the reference terms are the absolute contribution to the daily return. In this regard, we are not interested in other relative terms. We find that solar term 1 and 3 are robust at 5% level and 10% significance level respectively under EBA as they are always positive in both cases. Solar term 2 is nearly negatively robust at 90% level whereas solar term 4 and 13 are fragile under EBA. Therefore, we regard solar term 1 and 3 as somehow strongly positive terms. These solar terms are important findings in inter-solar term section. In the following sections, we shall examine them from a time series perspective combining with normal days. The fact is that, those solar terms we mention in Table 2 and Table 3 also show up frequently in the later sections.

Table 3   EBA result for inter-solar-term regression

| Solar Term | Estimated Coefficient (EBA error) | EBA Lower interval (95%) | EBA Upper interval (95%) | EBA Lower interval (90%) | EBA Upper interval (90%) |
|---|---|---|---|---|---|
| No.1 (Reference term) | 0.0086** (0.0029) | 0.0029 | 0.0168 | 0.0038 | 0.0133 |
| No.2 (Reference term) | -0.0050 (0.0039) | -0.0126 | 0.0026 | -0.0114 | 0.0001 |
| No.3 (Reference term) | 0.0124* (0.0070) | -0.0013 | 0.02612 | 0.0009 | 0.0239 |
| No.4 (Reference term) | -0.0083 (0.0084) | -0.0247 | 0.0082 | -0.0221 | 0.0054 |
| No.13 (Reference term) | 0.0065 (0.0056) | -0.0044 | 0.0174 | -0.0027 | 0.0157 |
| obs | | | 436 | | |

\* stands for 10% level significance under EBA robustness, \*\* stands for 5% level significance under EBA robustness.

*3.2. Full sample empirical result*

In this section, we analyze solar term effect through a time series perspective as well as involving with normal days (non-solar-term days). It is reasonable to say that the effect (anomaly) to daily return brought by solar terms may be time varying, influenced by the status of current period of return and some solar terms are efficient comparing with normal days even though they do not show significant contribution to return in inter-solar-term framework. Therefore, we add lagged value of daily return and return of normal days to our model and analyze solar term effect both under mean level (equation (7)) and risk level (equation (8)).

Table 5 shows the full result by applying equation(7). The constant term $\mu$ then represent the return to normal days and it is not significant which is consist with the feature of Shanghai Index in China (Table 4 and Figure 2). The overall return is almost zero with a minor positive value representing the long uptrend in the past decades. Return of Shanghai Index also shows a high peak and tick tail feature which is also in line with many existing literature in financial series study (Corlu, Meterelliyoz and Tiniç, 2016; Yan and Han, 2019).

**Table 4** Statistics of overall return of Shanghai Index

| Statistics | Result |
|---|---|
| Mean value | $4.1490 \times 10^{-4}$ |
| Standard deviation | 0.0172 |
| Skewness | 0.6840 |
| Kurtosis | 25.1800 |

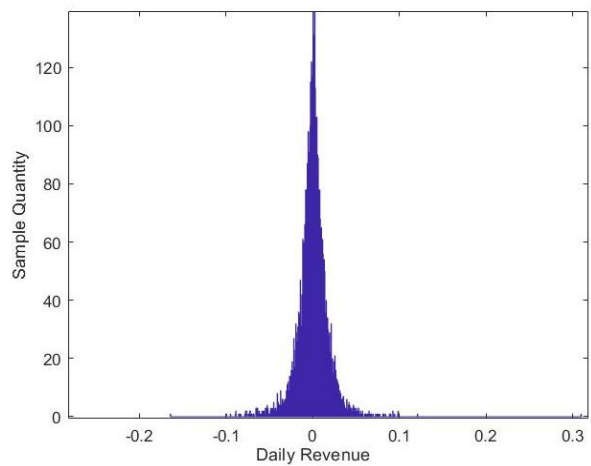

**Fig. 2.** Distribution plot of daily return (revenue) of Shanghai Index from 1995-2021

Among all solar terms, solar term 1, 3, 4 and 13 have significant effect to index return. Solar term 1, 3 and 13 are positive and solar term 4 is negative which is analogue to the result in Section 3.1. This reveals that these four solar terms are stable and outstanding enough both in inter-solar-term competition as well as comparing with normal days in full sample. We make a regression one more time only involving the four above (plus normal days and lagged return) in order to test the robustness alone. We put the result in Table 6 confirming the robustness successfully. Therefore, we claim that at mean level, solar term 1, 3 and 13 are positive efficient and solar term 4 is negative efficient. Investors can at least get profit at a significantly high probability through buying and selling within days or apply other index derivations. Note that this strategy only accounts for intraday behavior, combining with other technical and fundamental skills, investors may be confident to hold for a longer period.

**Table 5.** Result of full sample under mean level (equation (7))

| Coefficient | Estimation | Std Derivation | $t$ | $p$-value |
|---|---|---|---|---|
| $\mu$ | 0.0002 | 0.0002 | 1.1095 | 0.2673 |
| $r$ | 0.0306 | 0.0122 | 2.5189 | **0.0118**** |
| $\alpha_1$ | 0.0083 | 0.0038 | 2.2072 | **0.0273**** |
| $\alpha_2$ | -0.0051 | 0.0039 | -1.3122 | 0.1895 |
| $\alpha_3$ | 0.0124 | 0.0047 | 2.651 | **0.0080***** |
| $\alpha_4$ | -0.0089 | 0.0047 | -1.9182 | **0.0551*** |
| $\alpha_5$ | 0.0039 | 0.0038 | 1.038 | 0.2993 |
| $\alpha_6$ | 0.0052 | 0.0039 | 1.3502 | 0.1770 |
| $\alpha_7$ | -0.002 | 0.0056 | -0.3605 | 0.7185 |
| $\alpha_8$ | 0.002 | 0.0039 | 0.522 | 0.6017 |
| $\alpha_9$ | 0 | 0.0047 | 0.0088 | 0.9929 |
| $\alpha_{10}$ | 0.0031 | 0.0038 | 0.8253 | 0.4092 |
| $\alpha_{11}$ | 0.0038 | 0.0039 | 0.9835 | 0.3254 |
| $\alpha_{12}$ | 0.0006 | 0.0038 | 0.1653 | 0.8687 |
| $\alpha_{13}$ | 0.0063 | 0.0038 | 1.6863 | **0.0918*** |
| $\alpha_{14}$ | 0.0054 | 0.0038 | 1.4487 | 0.1475 |
| $\alpha_{15}$ | -0.0021 | 0.0039 | -0.5399 | 0.5893 |
| $\alpha_{16}$ | -0.0004 | 0.0038 | -0.1034 | 0.9177 |
| $\alpha_{17}$ | 0.0024 | 0.0039 | 0.6199 | 0.5354 |
| $\alpha_{18}$ | 0.0032 | 0.0039 | 0.8295 | 0.4068 |
| $\alpha_{19}$ | 0.0019 | 0.0040 | 0.4672 | 0.6403 |
| $\alpha_{20}$ | 0.0046 | 0.0038 | 1.2169 | 0.2237 |
| $\alpha_{21}$ | -0.004 | 0.0038 | -1.0569 | 0.2906 |
| $\alpha_{22}$ | 0.0005 | 0.0038 | 0.1438 | 0.8856 |
| $\alpha_{23}$ | 0.0013 | 0.0038 | 0.3553 | 0.7224 |
| $\alpha_{24}$ | 0.0024 | 0.0038 | 0.6448 | 0.5191 |
| obs | 6800 | 6800 | 6800 | 6800 |

* stands for 10% level significance, ** stands for 5% level significance, *** stands for 1% level significance.

Table 6. Result with refined solar terms under mean level (equation (7))

| Coefficient | Estimation | Std Derivation | t | p-value |
|---|---|---|---|---|
| $\mu$ | 0.0003 | 0.0002 | 1.5233 | 0.1277 |
| $r$ | 0.0318 | 0.0121 | 2.6215 | 0.0088*** |
| $\alpha_1$ | 0.0082 | 0.0038 | 2.1872 | 0.0288** |
| $\alpha_3$ | 0.0123 | 0.0047 | 2.6373 | 0.0084*** |
| $\alpha_4$ | -0.009 | 0.0047 | -1.9386 | 0.0526* |
| $\alpha_{13}$ | 0.0063 | 0.0038 | 1.6676 | 0.0955* |
| obs | 6800 | 6800 | 6800 | 6800 |

* stands for 10% level significance, ** stands for 5% level significance, *** stands for 1% level significance.

In addition, this paper also investigate solar term effect under risk level (i.e. volatility) as we know that China, as a emerging economy, has a unique stock market which is highly volatile in the past years. ARCH effect is verified in Table 7.

Table 7. Result of ARCH test in residual series

| Lagged order | Chi-squared | p-value |
|---|---|---|
| 1 | 170.8*** | 0.00 |
| 2 | 207.8*** | 0.00 |
| 3 | 516.7*** | 0.00 |
| 4 | 522.6*** | 0.00 |
| 5 | 524.9*** | 0.00 |
| 10 | 535.2*** | 0.00 |
| 15 | 5442*** | 0.00 |

*** stands for 1% level significance. The output is run by R program ArchTest in FinTS. Null hypothesis: No ARCH effect. The result shows that all lags test reject the null and therefore, ARCH effect is confirmed.

We apply an IGARCH (1,1) model to capture the volatility of daily return through equation (8). Worth noticing is that assumption for error distribution is sensitive to the result. Therefore, we set three assumptions: Normal distribution, Student-t distribution and Generalized error distribution which their assumptions are stricter one after one.. We confirm the existence of high volatility in daily return as two IGARCH terms are significant at 1% level. Under Normal distribution, solar term 1, 2, 3, 4, 5, 7, 8, 14 and 19 are significant to cause high volatility to the return and under Student-t distribution, the result only left solar term 1, 2, 4, 8, 14 and 19. When it comes to the strictest assumption: GED distribution, only solar term 1, 2, 4 and 14 are significant. With the strictness of assumptions increase, significant solar terms reduce. In summary, solar term 1, 2, 4 and 14 are the strongest (i.e. the most efficient terms at volatility level) and for the rest part, it depends. The result is meaningful to us as we know to which solar term we should pay attention based on our strictness to the error's distribution. Anyway, solar term anomaly in volatility level has been confirmed in this regard. Investors are supposed to take strategy like option volatility strategy to capture profit in volatile solar term days and the radicalness of the strategy should be considered according to the strongness of that solar term correspondingly.

Table 8. Result with refined solar terms under volatility level (equation (8))

| Coefficient | Assumptions of residual distribution | | |
|---|---|---|---|
| | Normal distribution | Student-$t$ distribution | Generalized error distr. |
| $\gamma$ | 0.051785*** (0.001745) [0.0000] | 0.060658*** (0.003789) [0.0000] | 0.059326*** (0.003542) [0.0000] |
| $\beta$ | 0.948215*** (0.001745) [0.0000] | 0.93934*** (0.00378) [0.0000] | 0.940674*** (0.003542) [0.0000] |
| $\alpha_1$ | 0.0000752*** 0.0000134) [0.0000] | 0.000067*** (0.000024) [0.0071] | 0.000071*** ()0.000025 [0.0061] |
| $\alpha_2$ | 0.0000949*** (0.0000157) [0.0000] | 0.000050* (0.000029) [0.0869] | 0.000053* (0.000028) [0.0652] |
| $\alpha_3$ | -0.0000849*** (0.0000156) [0.0000] | - | - |
| $\alpha_4$ | 0.0000834*** (0.0000205) [0.0000] | 0.000069* (0.000041) [0.0986] | 0.000054 (0.000039) [0.1720] |
| $\alpha_5$ | -0.0000521*** (0.0000115) [0.0000] | - | - |
| $\alpha_7$ | 0.000172*** (0.0000354) [0.0000] | - | - |
| $\alpha_8$ | 0.0000332*** <br> (0.0000154) <br> [0.0315] | 0.000048** <br> (0.000021) <br> [0.0280] | - |
| $\alpha_{14}$ | 0.0000423*** <br> (0.0000088) <br> [0.0000] | 0.000054** <br> (0.000023) <br> [0.0204] | 0.000051** <br> (0.000022) <br> [0.0192] |
| $\alpha_{19}$ | 0.0000379*** <br> (0.0000134) <br> [0.0048] | 0.000070** <br> (0.000027) <br> [0.0110] | - |
| Distribution-specific parameter | - | 4.747905*** <br> (0.215014) <br> [0.0000] | 1.159934*** <br> (0.015435) <br> [0.0000] |
| Obs | 6800 | 6800 | 6800 |

Standard errors are reported in parentheses and *p* values are reported in brackets.
* stands for 10% level significance, ** stands for 5% level significance, *** stands for 1% level significance.

### *3.3. Turn of solar term analysis*

In this section, we present the result of the turn of solar term effect as a supplementary to complete our study on solar term anomaly. Similar to the turn of week effect, we investigate the feature of daily return series from one day before to after a solar term. As this is a multiple days duration, it rather make sense to focus on duration volatility than return itself. Therefore, we make a slight modification to equation (8). In this case, $ST_{it} = 1$ refers that it is either one day before, after or simply on solar term *i*. We exclude none trading day so maybe some solar terms in certain year have less than 3 dummy variable that equals to 1.

The result is presented in Table 9. Still, we conduct the regression under three distribution assumptions as in the last section. Under Normal distribution: solar term 1, 2, 4, 5, 8, 9, 10, 11 and 14 are significant to induce high volatility. Solar term 8, 11 and 14 are significant under Student-t distribution whereas under GED distribution, there are solar term 2, 8, 11 and 14. These solar terms above are regarded that having a high volatility within their duration confirming the anomaly we expect. In this case, solar term 8, 11 and 14 are strongly efficient as they are significant in all cases.Likewise, we set the condition for 2 days before and after each solar term to extend the range. The result is presented in Table 10. In this case, solar term 8 and 11 are strongly efficient in causing high volatility. Finally, we conclude that solar term 8 and 11 are the strongest and robustest and solar term 14 is the second.

**Table 19.** Result of IGARCH(1,1) regression for turn of solar term effect (1-day range)

| Coefficient | Assumptions of residual distribution | | |
|---|---|---|---|
| | Normal distribution | Student-*t* distribution | Generalized error distr. |
| $\gamma$ | 0.061676*** | 0.064027*** | 0.065656*** |
| | (0.001928) | (0.003939) | (0.003962) |
| | [0.0000] | [0.0000] | [0.0000] |
| $\beta$ | 0.938324*** | 0.935973*** | 0.934344*** |
| | (0.001928) | (0.003939) | (0.003962) |
| | [0.0000] | [0.0000] | [0.0000] |
| $\alpha_1$ | 0.000021*** | | |
| | (0.000005) | | |
| | [0.0001] | | |
| $\alpha_2$ | 0.000031*** | | 0.000040*** |
| | (0.000005) | | (0.000011) |
| | [0.0000] | | [0.0003] |
| $\alpha_4$ | 0.000017*** | | |
| | (0.000007) | | |
| | [0.0179] | | |
| $\alpha_5$ | -0.000012** | | |
| | (0.000005) | | |
| | [0.0154] | | |
| $\alpha_8$ | 0.000042*** | 0.000022*** | 0.000024** |
| | (0.000007) | (0.000008) | (0.000009) |
| | [0.0000] | [0.0083] | [0.0117] |
| $\alpha_9$ | 0.000015 | | |
| | (0.000009) | | |
| | [0.1351] | | |
| $\alpha_{10}$ | -0.000015*** | | |
| | (0.000004) | | |
| | [0.0001] | | |
| $\alpha_{11}$ | 0.000020*** | 0.000014* | 0.000016* |
| | (0.000005) | (0.000008) | (0.00000) |
| | [0.0000] | [0.0784] | [0.0714] |
| $\alpha_{14}$ | 0.000016*** | 0.000017** | 0.000017** |
| | (0.000004) | (0.000008) | (0.00000) |
| | [0.0000] | [0.0272] | [0.0332] |
| Distribution-specific parameter | | 5.015936*** | 1.157359*** |
| | | (0.226108) | (0.016016) |
| | | [0.0000] | [0.0000] |
| Obs | 6800 | 6800 | 6800 |

Standard errors are reported in parentheses and *p* values are reported in brackets.
* stands for 10% level significance, ** stands for 5% level significance, *** stands for 1% level significance.

Table 10. Result of IGARCH(1,1) regression for turn of solar term effect (2-day range)

| Coefficient | Assumptions of residual distribution | | |
|---|---|---|---|
| | Normal distribution | Student-$t$ distribution | Generalized error distr. |
| $\gamma$ | 0.068789*** | 0.074649*** | 0.071584*** |
| | (0.003252) | (0.006829) | (0.006379) |
| | [0.0000] | [0.0000] | [0.0000] |
| $\beta$ | 0.928482*** | 0.920064*** | 0.922156*** |
| | (0.002929) | (0.006200) | (0.005915) |
| | [0.0000] | [0.0000] | [0.0000] |
| $\alpha_1$ | 0.000012** | | |
| | (0.000004) | | |
| | [0.0138] | | |
| $\alpha_2$ | 0.000021*** | | |
| | (0.000004) | | |
| | [0.0000] | | |
| $\alpha_4$ | 0.000011** | | |
| | (0.000005) | | |
| | [0.0434] | | |
| $\alpha_5$ | -0.000014*** | | |
| | (0.000004) | | |
| | [0.0012] | | |
| $\alpha_8$ | 0.000017*** | 0.000016* | 0.000015* |
| | (0.000005) | (0.000008 | (0.000009) |
| | [0.0000] | [0.0739] | [0.1000] |
| $\alpha_9$ | 0.000045*** | | |
| | (0.000006) | | |
| | [0.0000] | | |
| $\alpha_{10}$ | -0.000019*** | | |
| | (0.000003) | | |
| | [0.0000] | | |
| $\alpha_{11}$ | 0.000015*** | 0.000016** | 0.000017** |
| | (0.000004) | (0.000008) | (0.000008) |
| | [0.0000] | [0.0419] | [0.0364] |
| $\alpha_{14}$ | 0.000007** | | |
| | (0.000003) | | |
| | [0.0261] | | |
| $\alpha_{17}$ | 0.000006* | | |
| | (0.000004) | | |
| | [0.0786] | | |
| Distribution-specific parameter | | 4.512373*** | 1.151561*** |
| | | (0.261614) | (0.017797) |
| | | [0.0000] | [0.0000] |
| Obs | 6800 | 6800 | 6800 |

Standard errors are reported in parentheses and *p* values are reported in brackets.
* stands for 10% level significance, ** stands for 5% level significance, *** stands for 1% level significance.

## 4. Conclusions

This paper investigates the existing solar terms effect (anomaly) in China stock market based on Shanghai Index from 1995 to 2022. By applying a regression framework and EBA method, this paper mainly verifies the effect from three dimensions: inter-solar-term analysis, full sample analysis both at mean and risk level under multiple error distribution assumptions as well as the turn of solar terms effect. The result is remarkable: not all solar terms are efficient but we still find out a couple of them significant in all cases conforming the robustness and efficiency of some particular solar terms. This gives investors in China stock market a clue and reference to gain excess profit.

In inter-solar-term analysis, among 24 solar terms, solar term 1, 3 and 13 turn out to be significantly positive (at 5%, 5% and 10% level) to daily return whereas solar term 4 is significantly negative at 10% level. Solar term 2 is another infrequent one having a relatively high negative contribution to return but fail to pass robust test. Under EBA, only solar term 1 and 3 pass the robust test at 5% and 10% level. This shows that not all solar terms are efficient and the power of them is different. Eventually, solar term 1 and 3 turn out to be the strongest efficient, solar term 4 is efficient but less strong and solar term 2 is not efficient under inter-solar-term analysis but rather to be a potential candidate in further sections.

In full sample analysis, we use a AR (1)-IGARCH (1,1) model to extend the study involving with both solar term days and non-solar-term days (i.e. normal days) to further verify the universality of solar term effect. At mean level, solar term 1, 3, 4 and 13 are efficient to cause significant contribution to daily return. Still, solar term 1, 3 and 13 are positive and solar term 4 is negative which is exactly consistent with the findings in inter-solar-term analysis. At risk level (volatility), we run the IGARCH model under three different error distribution assumptions: Normal distribution, Student-t distribution and Generalized error distribution. Under Normal distribution, solar term 1, 2, 3, 4, 5, 7, 8, 14 and 19 are significant to cause high volatility to the return and under Student-t distribution, the result only left solar term 1, 2, 4, 8, 14 and 19. When it comes to the strictest assumption: GED distribution, only solar term 1, 2, 4 and 14 are significant. With the strictness of assumptions increase, significant solar terms reduce. The result is a bit different from the analysis before but still noticing some of the familiar solar terms such solar tern 1, 2, 3 and 4.

We also investigate the turn of solar terms effect at risk level based on the same model. We set the range for one day and two days respectively and finally verified that solar term 8 and 11 are the strongest and robustest and solar term 14 is the second to cause high volatility within their duration.

From multiple dimensions and multiple tests, we confirm the existence, robustness and reliability of solar term effect (anomaly) in China stock market. This help investors to gain excess profit in real trading as well as a good supplementary to understand investors' behavior academically.

## 5. Discussions

Despite solar term effect (anomaly) is a brand-new discovery in the market, we are still able to find some potential connections and reasonable logic between solar term effect and prevailing calender effects from our result in this paper.

Firstly, solar term effect is robust and reliable in multiple dimensions (i.e. inter-solar-term and full sample) which illustrate that it is a sort of "financial-independent" factor no matter comparing among themselves or comparing with normal days. In other words, efficient solar terms are always efficient in most cases and levels. Most of the significant solar terms above are in the first half of the year (i.e. having small orders) and especially gather on the first couple of terms in the year at mean level. At risk level, the result expends. Worth noticing is that the holiday duration of the Chinese lunar new year (i.e. the Spring festival) is 7 days, always right before, right after or simply overlapping solar term 3. Solar term 3 (see in Appendix A.) has a equal meaning to lunar new year as they all stand for a fresh beginning to the coming year and warm days. In some years, solar term 3 is just the last (first) trading days before (after) the lunar new year holiday which gives us a clue that significant positive effect on solar term 3 could be explained as another form of Chinese Lunar New Year (CLNY) effect or holiday effect in China which is analogue to the conclusions of McGuinness and Harris (2011). Besides, Yuan and Gupta (2014) report this phenomenon in other Asian market which imply that the effect may not only in China but also in other Asian countries having a profound influence by traditional Chinese culture.

As we conclude, the first several solar term show significant effect both at mean level and risk level which also consists with the "culture bonus hypothesis" proposed by Chen and Chien (2011). They suggest that under Chinese tradition, employees are rewarded with a generous bonus before Lunar New Year, most often paid in January, the period of time overlapping the first several solar terms. They point out that it is analogous to the concept of house money, enhances the propensity to bear increased levels of risk, which in turn stimulates the demand for higher risk securities, particularly in a market that is mainly dominated by individual investors, as in Taiwan n (also affected by Chinese culture even more than the mainland). This may also help to explain the high volatility and positive effect of solar term1 because solar term 1 is right after the international new year holiday. Future study around solar term 1 and its duration may help better reveal January effect in China.

Following cultural impact, solar term effect may be also well-explained from the perspective of investor mood. With the coming new year (i.e. solar term 1 and 2 after international new year and solar term 3 after lunar new year), people get more hopes and enthusiasm to wish for a better year. The findings in this paper affirmative the narrative of Teng and Yang (2018) who conclude that using emotion proxies provided by the literature, investors' good mood toward the festivities can help explain the CLNY (Chinese lunar new year) effect. While the effect remains strong for the long term, the CLNY effect gradually subsides after the opening up of A- and B-share markets. To the extent that foreign investors are less affected by the CLNY, the increased foreign participation in Chinese stock markets may help explain the diminishing CLNY effect. In addition, with the rise of temperature in the beginning of a year, investors' mood become active In this regard, solar term 13 and 14 are in the dog days in the year causing human fluctuation in mood. Our implication is in line with Kang et al. (2009) who find that the weather effect has a strong influence on the volatility of both A- and B-share returns in China market and the weather effect on volatility is explained by the openness of B-share market indicating that the market openness to domestic investors results in the weather effect. Liu (2013) also verifies the impact on investors' mood by weather and climate in a similar way.

Another interesting coincidence is that Wang, Lin and Chen (2010) figure out a lunar cycle effect. Their results imply that the moon affects individual mood and thinking, and lead to stock market change. They investigate new moon and full moon effect finding significant return in certain period. New moon falls usually on the first day of a lunar month and it is a crescent moon. Full moon is the 15th day of each lunar moth and new-full-new moon composes a month cycle in Chinese lunar calender. That's almost the analogue date of monthly solar term. Solar terms fall not necessarily the same day as new and full moon but the error is within one day. Solar terms have the same 15+15-day cycle as moon.

The solar term effect in this paper is a new perspective to view calender effect under the influence of traditional Chinese culture. In general, solar terms in particular period and seasons affect investors' mood and enthusiasm in trade and further form anomaly which shows an contradictory to the EMH making it possible for other studies and efforts in seeking regulations in stock market through price, volume and time (Sullivana, Timmermannb and Whiteb, 2001; Xiong et al., 2018).

**Conflict of interest**

There's no any conflict of interest in the article, all the work and study are approved and supported by the corresponding author and the institution. All the references and theories are appropriately cited in the article.

**Acknowledgment**

We are grateful for the instructions and help from Dr. Zhao as well as the continuous support from College of Science, Beijing Forestry University (BJFU).

**Appendix A: Introduction of Chinese twenty-four solar terms**

The Chinese twenty-four solar terms (known as "Jie Qi" in Chinese) is a very important part of ancient Chinese culture and guidance on agricultural activities even today (Chen, Li and Li, 2023).. In traditional Chinese culture, there are 360 days roughly in a year divided into 24 equal period which is led by each solar terms and there are around 15 days during each solar term period on average. Solar term comes one after another as a time duration and each solar term domains the following 14 days (i.e. altogether 15 days on average). Every solar term has a different name and a unique meaning showing the change and feature of the seasons at different levels. Today's twenty-four solar terms are based on the position of the sun on the ecliptic. That's the annual motion track of the sun which is divided into 24 equal parts, each 15° is one part as 1° equals one day and a year equals 360°. In other words, Chinese twenty-four solar terms can be regarded as 24 seasons in China instead of the common four seasons (Spring, Summer, Fall and Winter) which we are using internationally nowadays. The Chinese twenty-four solar terms is a fantastic refinement and summary of the change and transition of climate. It tells farmer when to sow, when to grow and when to harvest. Surprisingly, the climate always show significant fluctuation (e.g. obvious rise or fall of temperature, snowing, heavy rain, etc.) on the same day when a solar term falls. This kind of change of climate usually lasts in the rest days this solar term domains until the next solar term arrive. Therefore, temperature and climate change stage by stage according to the arrival of each solar term which compose the annual season (Qian, Yan and Fu, 2012). In addition, the 15-day duration of each solar term is further divided into three 5-day periods as three "Hou" in Chinese. The climate and temperature is thought to be changed more specificity at a 5-day period. Note that solar terms are no equal, 8 out of them are strong and important. Solar term 3, 9, 15 and 21 are the beginning of each season. Each season reaches its peak at solar term 6, 12, 18 and 24.

Nowadays, twenty-four solar term is mostly used in traditional Chinese medicine (TCM) and its relevant treatment such as acupuncture and so on. Solar terms have a non-negligible influence on human body, health and mood. Therefore, it is reasonable to say that solar term may have a indirect impact on stock market through affecting investors mood at some particular date. Finding out the potential influence on stock market would be definitely a big step forward of solar terms' value.

Table A.1   Meanings of solar terms

| Solar term order | Solar term name | Solar term date | Solar term meaning |
|---|---|---|---|
| 1 | Xiaohan (light snow) | Jan 5-Jan 7 | Happens right after new year with significant fall in temperature. |
| 2 | Dahan (Great snow) | Jan 20-Jan 21 | Happens before the Spring festival. Temperature reaches the trough. Coldest days. |
| 3 | Lichun (start of spring) | Feb 3-Feb 5 | End of winter, begins of spring and lunar new year. Ice begin melting |
| 4 | Yushui (water from rain) | Feb 18-Feb 20 | Significant increase in humidity, always with the first rain in spring. |
| 5 | Jingzhe (awakening of insects) | Mar 5-Mar 7 | Hibernation ends, revival of insect. |

| 8 | Gu'yu (rain to grain) | Apr 19-Apr 21 | Humidity steadily rise, time for initial period of farming. |
| 11 | Mangzhong （grain in ear) | Jun 5- Jun 7 | Busy and the best time in growing crops and grain. Early days in summer. |
| 13 | Xiaoshu (light heat) | Jul 6-Jul 8 | Rise in temperature. There will be dog days soon |
| 14 | Dashu (great heat) | Jul 22- Jul24 | Hottest days in a year, middle of dog days. High humidity. |
| 19 | Hanlu (cold dew) | Oct 8-Oct 9 | After National Day, Temperature begin to drop day by day. |